# Feasibility of Simulated Annealing Tomography

Nghia T. Vo,[1,*] Mark B. H. Breese,[2] and Herbert O. Moser[3]

[1] *Diamond Light Source, Harwell Science and Innovation Campus, Didcot, Oxfordshire, OX11 0DE, UK.*

[2] *Singapore Synchrotron Light Source, National University of Singapore, 5 Research Link, 117603, Singapore.*

[3] *Network of Excellent Retired Scientists and Institute of Microstructure Technology, Karlsruhe Institute of Technology, Postfach 3640, D-76021 Karlsruhe, Germany.*

*nghia.vo@diamond.ac.uk

**Abstract:** Simulated annealing tomography (SAT) is a simple iterative image reconstruction technique which can yield a superior reconstruction compared with filtered back-projection (FBP). However, the very high computational cost of iteratively calculating discrete Radon transform (DRT) has limited the feasibility of this technique. In this paper, we propose an approach based on the pre-calculated intersection lengths array (PILA) which helps to remove the step of computing DRT in the simulated annealing procedure and speed up SAT by over 300 times. The enhancement of convergence speed of the reconstruction process using the best of multiple-estimate (BoME) strategy is introduced. The performance of SAT under different conditions and in comparison with other methods is demonstrated by numerical experiments.

## I. INTRODUCTION

Simulated annealing (SA) is an algorithm that mimics the way a thermal system approaches its equilibrium state by slowly decreasing the temperature. Its principle can be used as a search engine to find a global minimum of optimization problems [1]. In the field of tomography, the SA algorithm has been used for image reconstruction in coded-aperture imaging [2]-[5], SPECT (single photon emission computed tomography) [6], PET (positron emission tomography) [7], or EIT (electrical impedance tomography) [8]. In transmission tomography, Haneishi *et al.* [9] used SA to reconstruct a 2D simulated blood vessel and investigated its performance under the modification of the cost function [10]. Qureshi *et al.* [11] compared the reconstruction results from SAT (simulated annealing tomography), FBP (filtered back-projection), and ART (algebraic reconstruction technique), which showed the superior quality of the SA-based method over the others. Moreover, he proposed various annealing schedules which are relevant to the complexity of objects. Although promising, SAT is a very time-consuming technique in which most of calculation time is used for determining the cost function which includes performing discrete Radon transform (DRT). Iteratively computing DRT is impractical for large size images. Here, we propose an approach which excludes the step of calculating DRT by using a pre-calculated intersection lengths array (PILA). This approach could speed up SAT more than 300 times compared with a conventional method including DRT. The significance of the PILA method is that it could be used for other iterative image reconstruction techniques where a set of projections (sinogram) of an image need to be recalculated for each change of the image. PILA is a technique which reduces the computational cost of the SA algorithm in the context of tomography, so it should not be confused with other techniques applicable for a general SA algorithm, such as modifying the sampling technique [12], determining the critical temperature [13], changing the cooling schedule [14], or combining with another search engine [15]. The improved SAT allows us to investigate its performance under different conditions where we found an interesting strategy, using the best of multiple estimates (BoME), which could enhance the convergence speed of SAT. Numerical experiments related to the change of BoME and the annealing schedule are investigated. The usefulness of PILA for another iterative tomographic method is demonstrated by applying it on SIRT (simultaneous iterative reconstructive technique). The comparison of reconstruction results from SAT, SIRT, and FBP are shown.

## II. METHODS

### A. Simulated annealing tomography

In computerized tomography, a digitized 2D function or image (pixel unit) is reconstructed from a set of its 1D projections at different angles. Fig. 1 illustrates a phantom image and one of its projections which is given by the summation along rays inclined at an angle $\alpha$, or DRT

$$P(\alpha, k) = \sum_{k} f_k(i,j) l_k(i,j) \qquad (1)$$

where $f_k(i,j)$ is the grayscale of a pixel $(i,j)$ intersecting with the $k^{th}$ ray and $l_k(i,j)$ is the intersection length.

One difference to the FBP reconstruction, which performs the inverse process (starting from a set of 1D projections to reconstruct a 2D image), is that SAT is a forward method in which one start from an initial estimated image and iteratively change its gray-levels for its 1D calculated projections (given by performing DRT) to fit best the measured projections. The process of altering the estimated image is the simulation of an annealing process.

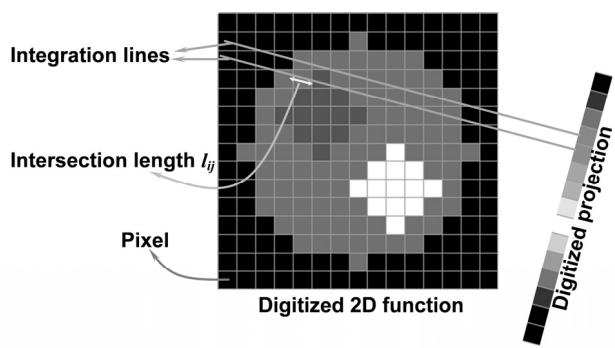

Fig. 1. Demonstration of a digitized 2D function and one of its projections. Because a pixel has a specific size, the calculation of a line integral through this function must include weight factors which are the intersection lengths between the ray and pixels.

The step-by-step analysis of SAT is as follows:

Step 1 (configuration step): The initial estimate could be chosen as a uniform gray image, a random grayscale image or a FBP-reconstructed image. The spatial constraint (the area of image where its values are different from zero) should be used to reduce calculation time if possible. A schedule of cooling down "temperature" (a parameter controlling the probability of accepting the estimated image) with initial temperature $T_0$ and final temperature $T_N$ is pre-set such as the linear form [11]

$$T_n = T_0 - n\frac{T_0 - T_N}{N} \qquad (2)$$

where $T_n$ is the temperature at the $n^{th}$ iteration and $N$ is the number of iterations which is preselected to terminate the calculation.

Step 2 (generation step): First, a pixel of the estimated image is chosen randomly. For generating the next value of the pixel, its range of values is limited to within $[0; \beta]$ where $\beta$ can be determined by using the maximum value of the FBP-reconstructed image or the back-projection image. The value of the pixel is updated as

$$f(i,j) = f(i,j) + random[-\beta; \beta] \qquad (3)$$

where $f(i, j)$ is the grayscale of the pixel at the position $(i, j)$, the $random[-\beta; \beta]$ function generates a random real number distributed uniformly in the range $[-\beta; \beta]$. The updating process is repeated if the new value of the pixel falls out of the range $[0; \beta]$.

Step 3 (evaluation step): The cost function is given by

$$C = \sqrt{\frac{1}{KU}\sum_{u=1}^{U}\sum_{k=1}^{K}\left[P_c(\alpha_u, k) - P_m(\alpha_u, k)\right]^2} \qquad (4)$$

where $U$ is the number of angles or projections, $K$ is the pixel number of each projection, $P_c(\alpha_u, k)$ is the calculated projection of the updated image at angle $\alpha_u$ and position $k$, and $P_m(\alpha_u, k)$ is the measured projection at the same angle and position. The calculated projections of the estimated image are given by (1) at every angle corresponding to the measured projections. This is the most time-consuming step of SAT. In practice, a tomographic data set can include millions of pixels and thousands of projections, hence applying DRT for every gray-level update of a single pixel is not a realistic task.

Step 4 (acceptance step): Calculate $\Delta C$ which is the difference of the cost function after and before the update. If $\Delta C<0$, the new value of the pixel is accepted, the calculated projections and the cost function is updated. If $\Delta C>0$, a random real number, $N_1$, in the range $[0;1]$ is generated and compared with the value $N_2=\exp(-\Delta C/T_n)$. If $N_1<N_2$, the new value of the pixel is accepted, the calculated projections and the cost function is updated. If $N_1>N_2$, we return to step 2. The meaning of this step is to avoid the probability of sticking in local minima.

The routine from step 2 to step 4 is under the control of the cooling schedule in which we can repeat the routine *M* times for each stage (temperature not changed) to increase the probability of converging to the global minimum. After *M* iterations for each temperature, the temperature is reduced by (2) and the process stops at the final temperature.

### B. Different approach for determining the cost function

The applicability of SAT is limited by the time consumption for calculating DRT (1) to determine the cost function. This consumption increases with the increase of the size of the image. To reduce the computing time, we propose an approach which excludes DRT from the determination of the cost function. The procedure is as follows:

First, we calculate a set of intersection lengths of different rays at different angles (equivalent to the angles of measured projections) by using Siddon's algorithm [16] for a single pixel *(i, j)* as demonstrated in Fig. 1. The stored information is formatted by

$$\{(i,j)\} = \{\{k_1, l_1\}, \{k_2, l_2\}, ..., \{k_u, l_u\}, ..., \{k_U, l_U\}\} \qquad (5)$$

where *{k_u,l_u}* respectively is the position on the projection (at angle $\alpha_u$) which gets the contribution from the pixel *(i,j)*, and the intersection length of the pixel *(i,j)* with the ray from that projection. The process is applied for all of the pixels of the image. The information is stored in the computer memory and recalled during the calculation. For saving RAM memory space, we convert the intersection length which is a real number staying in the range $[0;\sqrt{2}]$ (pixel size normalized to 1) to the 1-byte integer number by multiplying it with a factor of 180 and rounding the result.

Suppose that we start the calculation process with the initial calculated projections (using Siddon's algorithm [16]) from the initial estimate of the image. At the next step, the pixel *(i,j)* is chosen randomly and changed by an amount of *Δf*. From the stored information, we know that the change of the pixel *(i,j)* leads to the change of amount *Δf×l_u* at the position $k_u$ of the projection (at angle $\alpha_u$). Hence, instead of applying DRT (1) on the next estimated image to obtain the calculated projections, we only need to update the initial calculated projections at the positions and the amount of values corresponding to the pixel *(i,j)*. This approach is a huge time saving compared with the conventional way. We refer to this proposed technique as the pre-calculated intersection lengths array approach.

Substituting the updated value of calculated projections

$$P'_c(\alpha_u, k_u) = P_c(\alpha_u, k_u) + \Delta f \times l_u \qquad (6)$$

into (4) yields

$$C' = \sqrt{C^2 + \frac{1}{KU}\sum_{u=1}^{U}\left\{\begin{array}{l}2[P_c(\alpha_u,k_u) - P_m(\alpha_u,k_u)] \times \Delta f \times l_u \\ +(\Delta f \times l_u)^2\end{array}\right\}}. \tag{7}$$

which is the update of the cost function. This is another time-saving step in that we only need to add an amount of value to the previous cost function to obtain its update instead of using (4) for the whole projection data set. The form of (7) is somehow similar to the idea of "Delta function" [17] for solving timetabling problem. However, the Delta function is only applicable for discrete problems where the cost function has a finite number of values. Note that equation (7) is the consequence of the PILA method, not the main achievement of the paper. PILA is a general technique for quick updating the change of sinogram relevant to the change of the image without performing DRT. Its principle could be used for other iterative image reconstruction, demonstrated in section III-D.

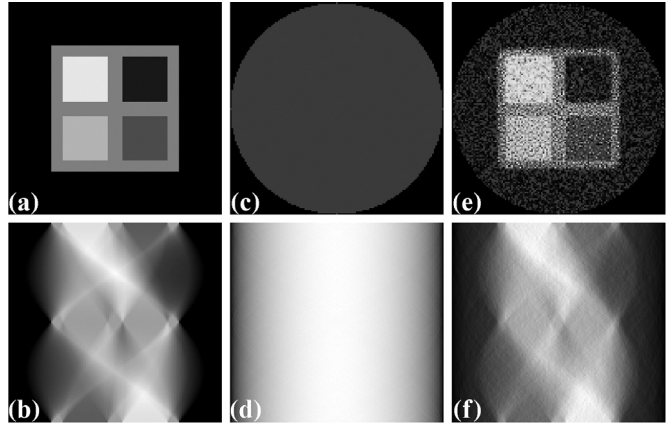

Fig. 2. Phantom (a) used for simulation and its sinogram (b). The initial estimate (c) and its sinogram (d). The estimated image in the reconstruction process (e) and its sinogram (f).

## III. SIMULATION RESULTS

In the following experiments we use some common initial set-ups including: the initial estimate is a uniform solid circle image as demonstrated in Fig. 2(c); the initial temperature is determined by choosing the probability about 90% for accepting estimates which return the increasing cost functions [7]

$$T_0 = -\frac{\sum_{q=1}^{Q}(C'-C)_q}{Q\ln(0.9)} \tag{8}$$

where Q is the number of estimates (a user-selected number, here we use *Q=200*), all generated from the same initial estimate, which give *C'>C*; the final temperature $T_N = T_0/10^3$; the cooling schedule follows (2); and *N=1000*. Other parameters are flexibly chosen depending on the size of the phantom. All experiments included here are performed in Mathematica 8.0 using the compile mode [18].

*A. Gain in speed*

For demonstrating the efficiency of the SAT based on the PILA approach, a performance comparison with the SAT included DRT using a bilinear interpolation scheme, which is the well-known standard method for computing projections [19], is performed using the same number of projections and iterations ($N=1$, $M=200$). The purpose of the experiment is to evaluate speed only, so we do not implement a full reconstruction. The results of the comparisons for different sizes of the phantom (Fig. 2(a)) at different number of projections ($U=90$, $U=180$, and $U=360$) are shown in Fig. 3. We see that the gain in speed increases with the increase of the image size and the number of projections. This improvement is very significant to iterative reconstruction methods in which the calculation time is counted by hours instead of seconds as with direct reconstruction methods.

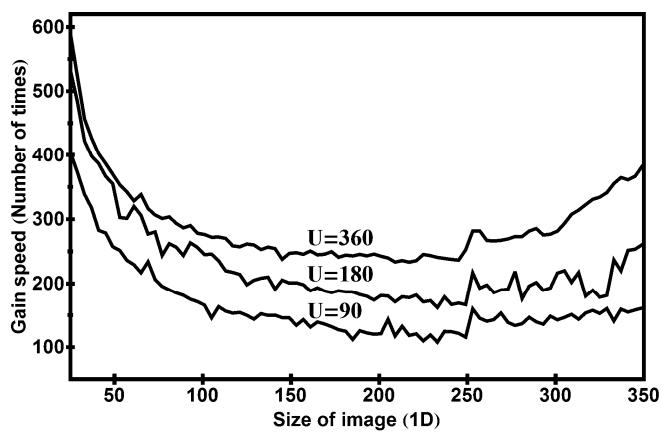

Fig. 3. Gain in speed of the SAT-PILA compared with SAT-DRT versus the size of the image.

*B. Using the best of multiple-estimate to increase the convergence speed*

Each estimate includes updating the change of: the estimated image, its sinogram, and its cost function. Suppose that we start from the estimated image $f^m$ at the $m^{th}$ iteration with its sinogram $P^m$ and cost function $C^m$. The principle of the BoME strategy is that, instead of generating only one estimate in the generation step of SAT we generate $N_e$ estimates and choose the one giving the smallest cost function. For example, we generate $N_e=3$ estimates $\{\{f^{m1}, P^{m1}, C^{m1}\}, \{f^{m2}, P^{m2}, C^{m2}\}, \{f^{m3}, P^{m3}, C^{m3}\}\}$. If $C^{m2}$ is the smallest value compared with $C^{m1}$ and $C^{m3}$, the next update is $\{f^{m+1}, P^{m+1}, C^{m+1}\}=\{f^{m2}, P^{m2}, C^{m2}\}$. Keeping in mind that although $N_e$ estimates has been made, there is only one update of the image, its sinogram, and its cost function. In addition, there is no relation between different sampling of $N_e$ estimate in which every pixel is chosen randomly in the reconstruction area (Fig. 2(c)). The numerical experiments are investigated under the conditions: $N=1000$, $M=2000$, $U=60$, $\beta=0.9$, image size $K \times K = 64 \times 64$ for the cases of $N_e=1, 2, 4, 8, 16$, and $32$ estimates. Fig. 4 shows reconstructed images for different $N_e$ and plots of the normalized root mean square error (NRMSE), which is the root mean square of the difference between the reconstructed and original image normalized by dividing the mean value of the original image, versus the number of iterations ($N$) for all investigated cases.

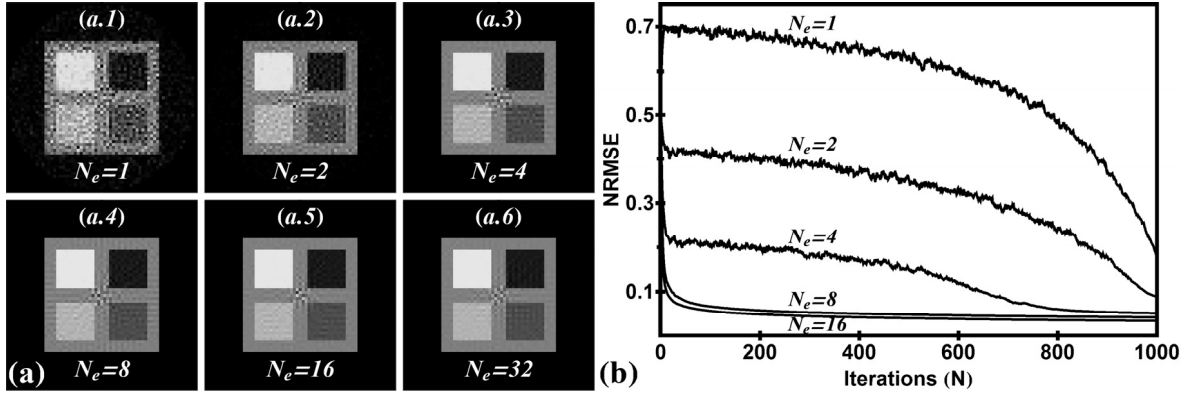

Fig. 4. Reconstructed images at different $N_e$ (a) and the comparison of the convergence speed (b) where the plot of $N_e=32$ is not shown due to it being nearly identical to that of $N_e=16$.

As can be seen in Fig. 4(b), the speed of convergence increases as the number of generated estimates increases. For instance, in the case of $N_e=16$, the reconstructed image reaches an error of less than 5% after only $N=200$ iterations. To show the advantage of using high values of $N_e$, we compare the convergence speed between the case of $N_e=4$ and the usual case, $N_e=1$, having the same number of calculations. Particularly, in the case of $N_e=4$, the total number of calculations or evaluations is $N \times M \times N_e = 8 \times 10^6$ (but there are only $N \times M = 2 \times 10^6$ times of updating the estimated image). Thus, to compare equivalently in the case of $N_e=1$ we increase the value of $N$ and $M$ for $N \times M = 8 \times 10^6$.

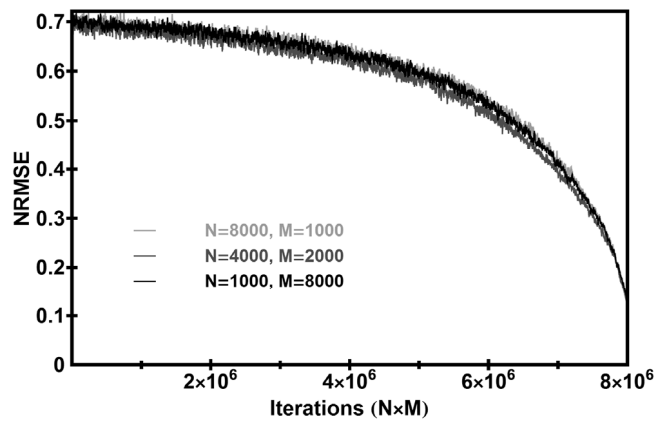

Fig. 5. Convergence process of the case of $N_e=1$ having the same number of evaluations as $N_e=4$.

As shown in Fig. 5, the final NRMSE of three combinations of N and M (of $N_e=1$) is still higher than the one of $N_e=4$ (Fig. 4(b)) although the number of updates of the estimated image is four times higher. However, there is no significant gain in convergence speed when $N_e$ reaches a certain value that depends on the size, but not on the complexity of the phantom as indicated in Fig. 6(c) and 6(d). These plots present the final cost function after $N \times M$ iterations versus number of estimates of different experiments with different sizes of phantom 1 (Fig. 2(a)), and different types of phantom (Fig. 6(a) and 6(b)). The conditions of these experiments are: $N=200$, M equals the number of pixels in the reconstruction space ($\sim \pi K^2/4$), $U=K$. From the results of Fig. 6(c) and 6(d) we see that $N_e$ can be chosen in the range of 8 to16 for a good convergence speed. The principle of BoME enables it to be used for parallel computing on multicore computers. It is different to other parallel techniques proposed for the SA algorithm which require communication between processors on the sampling distribution, cost function calculation, or synchronization acceptance [20]. There is no communication requirement between estimates on different processors in the BoME strategy.

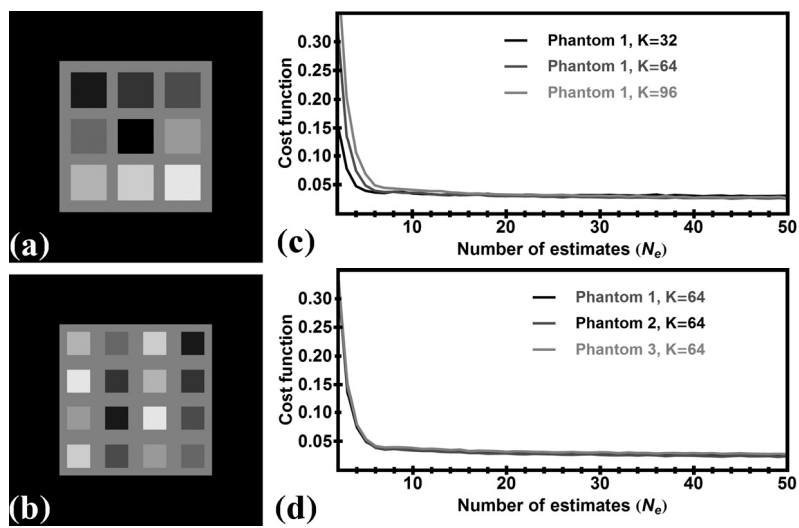

Fig. 6. Phantom 2 (a) and 3 (b) used for experiments. (c) is the plots of the cost function at the end of each experiment versus number of estimates for different sizes of phantom 1, (d) is the one for different kinds of phantom.

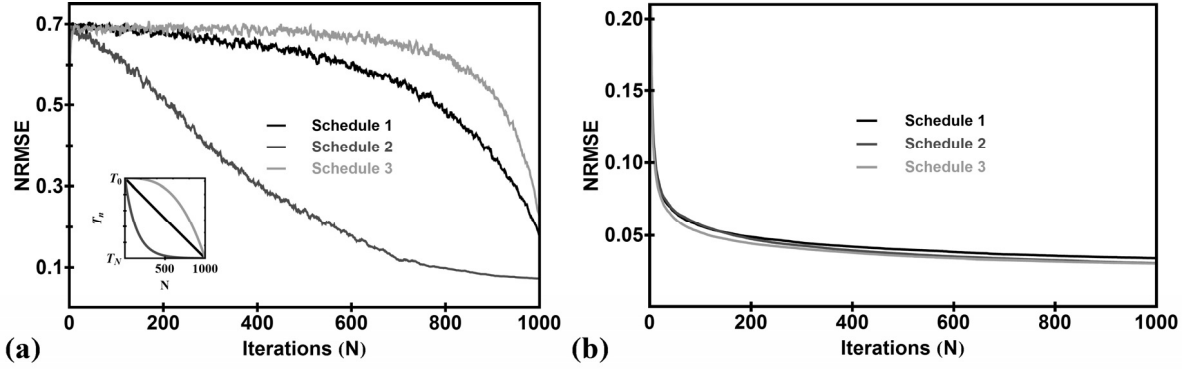

Fig. 7. Comparison of the performance of three schedules in the case of $N_e$=1 (a), and $N_e$=16 (b) where the small graph in (a) shows the plots of different cooling schedules.

*C. Impact of cooling schedule*

To test the impact of the cooling schedule, the performance of the schedule of (2) is compared with other schedules including: one which drops temperature faster [11]

$$T_n = T_0 \left(T_N/T_0\right)^{n/N}, \tag{8}$$

and one which drops temperature slower

$$T_n = T_0 - n^3 \frac{(T_0 - T_N)}{N^3}. \tag{9}$$

The conditions of the numerical experiments are the same as in section III-B for the two cases: $N_e$=1 and $N_e$=16. The results are shown in Fig. 7 where schedules 1, 2, and 3 represent formula (2), (8), and (9), respectively. In the case of $N_e$=1 (Fig. 7(a)), the impact of cooling schedules on the convergence speed is clear where the schedule having a faster temperature drop gives a better convergence speed. However, this effect does not show in the case of $N_e$=16 (Fig. 7(b)) where the convergence speed of schedule 3 is even slightly faster than the others. In practice, instead of looking for an optimum cooling schedule which is not always feasible and depends on the complexity of the phantom, we simply use the BoME strategy.

*D. Comparison with other methods*

SIRT is another well-known iterative reconstruction which uses intersection lengths in its formulae [21]. However, in practice, these values are simplified by rounding to 1 or 0 for saving computational cost. This simplification reduces the quality of the reconstructed image. Here, we apply the PILA technique to use exact values of intersection lengths in SIRT's formulae to improve the reconstruction result. Starting from the initial estimate as in Fig. 2(c), the gray level of a pixel is updated as

$$f(i,j) = f(i,j) + \sum_{u=1}^{U} \Delta f_u(i,j) \Big/ U \tag{10}$$

where

$$\Delta f_u(i,j) = \frac{P_m(\alpha_u,k) - P_c(\alpha_u,k)}{\sum l_{uk}^2} l_u. \tag{11}$$

$\sum l_{uk}^2$ is the summation of the square of intersection lengths of the $k^{th}$ ray at angle $\alpha_u$ through the reconstruction space (Fig. 2(c)). This factor can be pre-calculated by using Siddon's algorithm where the input is the binary image of Fig. 2(c), the intersection lengths is squared, and the output is a factor matrix having the same size as the sinogram. The iterative process of SIRT includes: firstly, a pixel in the reconstruction space is chosen. From PILA and the factor matrix we obtain information to compute (11). After updating the pixel, we use PILA again to update its sinogram for the next calculation step. The process is repeated $N$ times where each time $M$ pixels in the reconstruction space are updated.

Fig. 8 shows the reconstruction results from SIRT and SAT in the conditions of noise-free data and added SNR=50 white noise data [22]. Although the convergence speed of the cost function of SAT in both conditions (Fig. 8(f) and 8(g)) is faster than SIRT, the SIRT method returns better reconstructed images having errors of 2.9% and 14.6% in the cases of free-noise and noisy data, respectively, compared with 5.2% and 21.8% from the SAT reconstruction (Fig. 8(h) and (i)). However, we see artifacts in the reconstructed image of SIRT (Fig. 8(d)) at the free-space area of the phantom. These artifacts are nearly not shown in the result of SAT (Fig. 8(e)). In these experiments, the phantom (Fig. 8(a)) having a size *K=128* and its sinogram with *U=120* are generated. *N=1000* and *M≈πK²/4* are used for both SIRT and SAT where the extra parameters of SAT are *N_e=16* and *β=0.9*. The reconstructions of SIRT and SAT from noisy data can be improved by using regularization or smoothing techniques [23], which is out of the scope of this paper.

In the FBP method, for highly accurate reconstruction of an image $K \times K$, the number of projections $U$ is required to be approximate to $K$, *U ≈ (π/2)×K* [21]. This principle does not change even in many cases where the area of the constrained space (where the gray levels of the phantom are different from zero) is smaller than the reconstruction area. The advantage of SAT in these cases is that we can use a smaller number of projections than the FBP method, but still obtain the same reconstruction quality. For comparison between two methods on the performance related to the number of projections, the phantom in Fig. 8(a) is re-used, but its center stays out of the center of rotation.

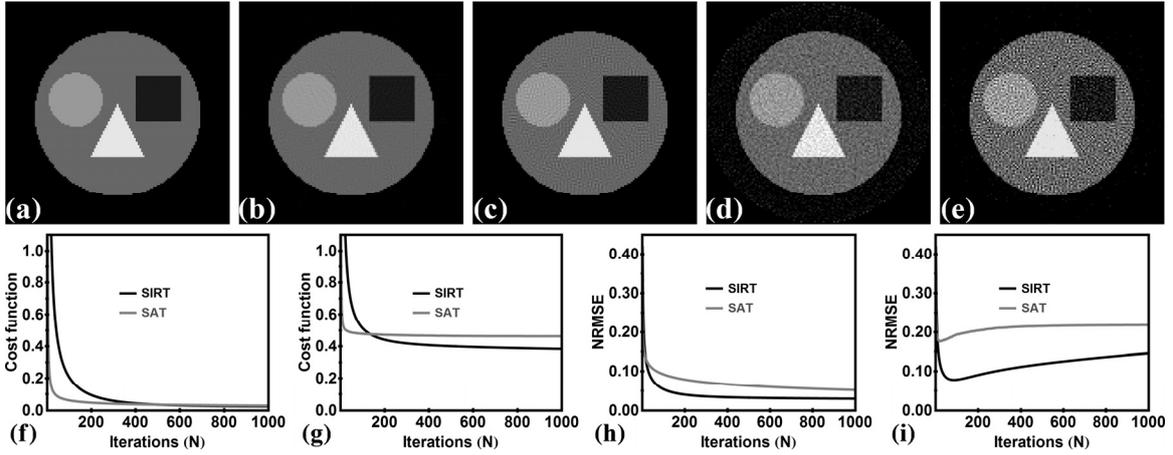

Fig. 8. Comparison between SIRT and SAT: (a) Phantom. (b) and (c) are its reconstructed images from SIRT and SAT in the case of noise-free data. (d) and (e) are the results of SIRT and SAT, respectively, from noisy data. The plots show the cost function and NRMSE versus $N$ iterations of both SIRT and SAT in the case of noise-free data ((f) and (h)), and noisy data ((g) and (i)).

The purpose of this change is to emphasize the disadvantage of FBP which returns a larger error of the reconstruction area farther to the center of rotation. This disadvantage shows more clearly when the number of projections is not large enough. Fig. 9(a) and 9(b) present the reconstruction results of FBP for two cases of $U=90$ and $U \approx (\pi/2) \times K \approx 200$ where the streak artifacts are visible in the case of lower U. Using the same number of projections, SAT gives a better reconstruction (Fig. 9(c)) if the spatial constraint is used, which means that in the generation step of the SA algorithm the next pixel is chosen randomly in the constrained space instead of a full reconstruction area. In practice, we should use the FBP-reconstructed image for generating the initial estimate and the constrained space (using threshold filter), which helps to improve the reconstruction quality and reduce the computational cost.

IV. CONCLUSION

In summary, we proposed the PILA approach which enables the fast update of the sinogram of an image relevant to the change of its pixels. This technique massively reduces the cost of iteratively computing DRT in the SAT method, which improves its feasibility. The SAT-improved method allows us to investigate its performance under different conditions where the finding of BoME strategy is very useful to enhance the convergence speed of image reconstruction. The usefulness of the PILA technique is also demonstrated on the SIRT method which shows the potential of applying SIRT-PILA for superior image reconstruction. The advantage of SAT over FBP is shown. The further improvements of SAT by using other optimum techniques developed for SA algorithm are interesting for further study.

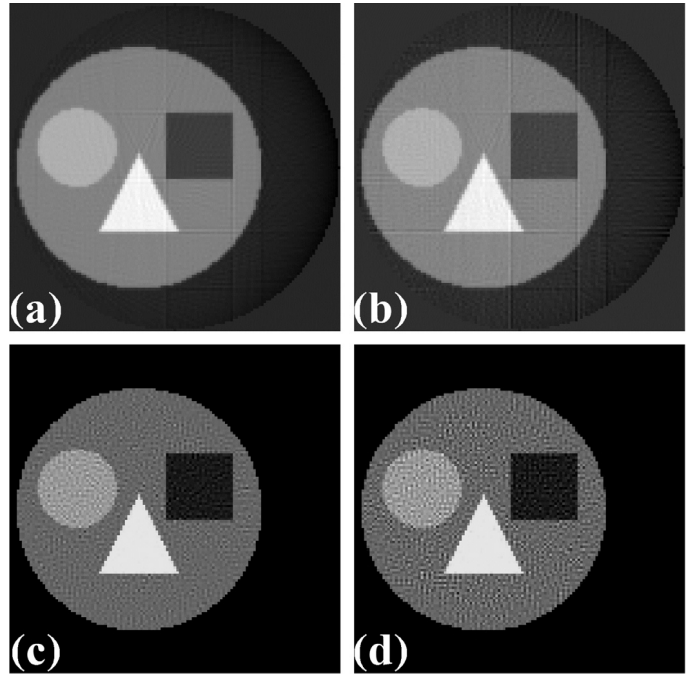

Fig. 9 The reconstruction from (a) FBP, *U=200*, *NRMSE=8.2%*, (b) FBP, *U=90*, *NRMSE=9%*, (c) SAT, *U=90*, with spatial constraint, *NRMSE=6.4%*, (d) SAT, *U=90*, without spatial constraint, *NRMSE=11.3%*.